\documentclass[prl,twocolumn,showpacs,showkeys,amsmath,amssymb]{revtex4}
\usepackage{graphicx}
\usepackage{hyperref}

\newcommand{\ket}[1]{\ensuremath{|#1\rangle}}

\begin{document}

\title{Transformed Dissipation in Superconducting Quantum Circuits}

\author{Matthew Neeley}
\affiliation{Department of Physics, University of California at Santa Barbara, Broida Hall, Santa Barbara, CA 93106}
\author{M. Ansmann}
\affiliation{Department of Physics, University of California at Santa Barbara, Broida Hall, Santa Barbara, CA 93106}
\author{Radoslaw C. Bialczak}
\affiliation{Department of Physics, University of California at Santa Barbara, Broida Hall, Santa Barbara, CA 93106}
\author{M. Hofheinz}
\affiliation{Department of Physics, University of California at Santa Barbara, Broida Hall, Santa Barbara, CA 93106}
\author{N. Katz}
\affiliation{Department of Physics, University of California at Santa Barbara, Broida Hall, Santa Barbara, CA 93106}
\altaffiliation{Department of Physics, Hebrew University, Jerusalem, Israel}
\author{Erik Lucero}
\affiliation{Department of Physics, University of California at Santa Barbara, Broida Hall, Santa Barbara, CA 93106}
\author{A. O'Connell}
\affiliation{Department of Physics, University of California at Santa Barbara, Broida Hall, Santa Barbara, CA 93106}
\author{H. Wang}
\affiliation{Department of Physics, University of California at Santa Barbara, Broida Hall, Santa Barbara, CA 93106}
\author{A. N. Cleland}
\affiliation{Department of Physics, University of California at Santa Barbara, Broida Hall, Santa Barbara, CA 93106}
\author{John M. Martinis}
\email{martinis@physics.ucsb.edu}
\affiliation{Department of Physics, University of California at Santa Barbara, Broida Hall, Santa Barbara, CA 93106}


\date{\today}

\begin{abstract}
Superconducting quantum circuits must be designed carefully to avoid dissipation from coupling to external control circuitry.  Here we introduce the concept of current transformation to quantify coupling to the environment.  We test this theory with an experimentally-determined impedance transformation of $\sim\,10^5$ and find quantitative agreement better than a factor of 2 between this transformation and the reduced lifetime of a phase qubit coupled to a tunable transformer.  Higher-order corrections from quantum fluctuations are also calculated with this theory, but found not to limit the qubit lifetime.  We also illustrate how this simple connection between current and impedance transformation can be used to rule out dissipation sources in experimental qubit systems.
\end{abstract}

\maketitle

The quantum behavior of superconducting circuits has been demonstrated by numerous experiments \cite{Sillanpaa2007,Majer2007,Plantenberg2007,Niskanen2007,Steffen2006}, and their promise as quantum information processors \cite{Nielsen2000,DiVincenzo2000} is well-established.  These devices must be carefully engineered to protect their quantum states from environmental noise, particularly that from control circuitry to which the qubits are permanently wired.  Environmental dissipation can generally be described by spin-boson models \cite{Makhlin2001,Wal2003} or---more practically---computed as being proportional to the real part of the admittance $Y(\omega)$, the classical response of the circuit \cite{Esteve1986}.

In this letter, we give the first description of how current transformation allows external sources of dissipation to be simply and physically determined.  This idea enables a direct comparison between measurements of energy decay $T_1$ and the transformed dissipation from the environment.  While our circuit uses an electrically tunable transformer similar to that demonstrated previously in superconducting circuits \cite{Bertet2005, Yoshihara2006}, the direct measurement of the current transformation allows a quantitative comparison to this theory, with agreement better than a factor of two.  This concept is presented in a general two-port model that has been extended beyond that of a classical impedance transformation to include effects of quantum fluctuations.  Although this work does not directly show how to improve $T_1$, currently an important issue, we illustrate how this general theory may be used to experimentally rule out sources of decoherence.

\begin{figure}[bp]
\begin{center}
\includegraphics[trim=0.2in 2.4in 0.2in 0.2in] {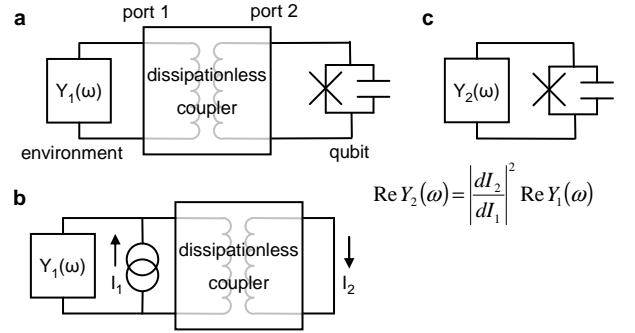}
\end{center}
\caption{Impedance transformation by a 2-port dissipationless coupler.  (\textbf{a}) Qubit is connected through a dissipationless coupler to environment, described by admittance $Y_1(\omega)$ of the external circuitry.  (\textbf{b}) The transfer function of the coupler relates current source $I_1$ across the input port to current $I_2$ across the shorted output.  (\textbf{c}) The effective dissipation seen by the qubit at port 2 is transformed by the squared derivative of the transfer function.}
\label{fig:BlackBox}
\end{figure}

A superconducting qubit is generally coupled to control circuitry via a dissipationless element, typically a capacitor or a mutual inductance.  As shown schematically in Fig. \ref{fig:BlackBox}, the external circuitry is characterized by its admittance $Y_1(\omega)$ (often $1/(50\,\Omega)$ from a transmission line), and the coupler transforms this into an effective admittance $Y_2(\omega)$ seen by the qubit.  From the fluctuation-dissipation theorem, one finds that the real part of the effective admittance seen at the output of the coupler is
\begin{equation}
\textrm{Re}\, Y_2(\omega) = \left|dI_2/dI_1\right|^2 \textrm{Re}\, Y_1(\omega) \ ,
\end{equation}
where $I_1$ is a current source applied at the input port, and $I_2$ is the resulting current that appears across the shorted output port (see Fig. \ref{fig:BlackBox}b). This admittance leads to a qubit lifetime \cite{Esteve1986} that is approximately equal to the classical decay time $T_1 \approx C/\textrm{Re}\,Y_2(\omega_{10})$, where $\omega_{10}$ is the qubit transition frequency, and $C$ is the qubit capacitance.  Thus, given an environment $Y_1(\omega)$, the current transfer function $I_2(I_1)$ completely determines the dissipation seen by the qubit through the coupler.  When fluctuations in $I_1$ produce no current $I_2$ in the qubit, the environment is decoupled. This simple result holds not only for capacitors and inductors, but also for more complicated nonlinear dissipationless couplers.

\begin{figure}[t]
\begin{center}
\includegraphics[trim=0 1.1in 0 0] {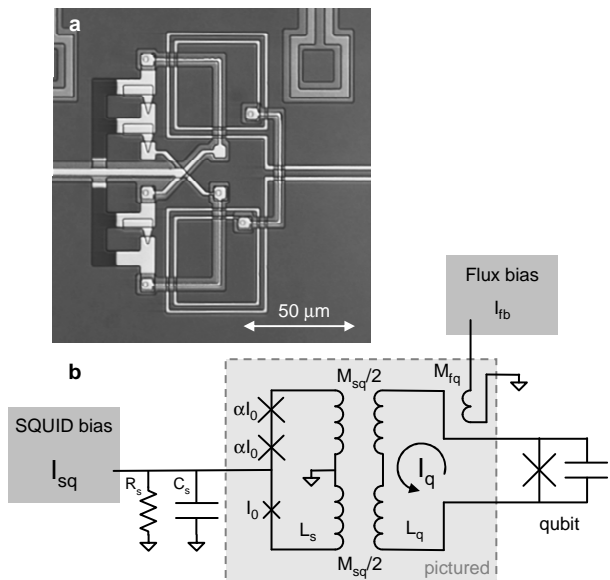}
\end{center}
\caption{Micrograph (\textbf{a}) and schematic (\textbf{b}) of the phase qubit and SQUID.  The overlap of the qubit and SQUID loops increases their mutual inductance $M_{sq}$, while their gradiometric layout reduces their sensitivity to external flux.  The flux bias coil couples to the qubit to tune its frequency, but has negligible mutual inductance with the SQUID.  The shunt resistor $R_s$ reduces quasiparticle generation in the SQUID when it switches \cite{Lang2003}.  The qubit sees $R_s$ transformed by coupling through the SQUID.  For the tested device, qubit capacitance and critical current are $1\,\textrm{pF}$ and $2\,\mu\textrm{A}$.  In addition, we have $I_0 = 2\,\mu\textrm{A}$, $\alpha = 1.5$, $L_s = 300\,\textrm{pH}$, $L_q = 720\,\textrm{pH}$, $M_{sq} = 70\,\textrm{pH}$, $M_{fq} = 2\,\textrm{pH}$, $R_s = 30\,\Omega$ and $C_s = 1\,\textrm{pF}$.}
\label{fig:Layout}
\end{figure}

To test this idea, we measured the lifetime of a flux-biased Josephson phase qubit as a function of the current bias through a 3-junction measurement SQUID \cite{Simmonds2004}.  The layout and schematic of the phase qubit and measurement SQUID is shown in Fig. \ref{fig:Layout}.  The operation of this device has been described previously \cite{Steffen2006a}, and we repeat here only the relevant details.  The qubit frequency is tunable over a range of several GHz by applying magnetic flux to the qubit loop.  The qubit state is measured by selectively tunneling the qubit $\ket{1}$ state out of the cubic well of the phase qubit potential.  The tunneled $\ket{1}$ and non-tunneled $\ket{0}$ states produce different amounts of magnetic flux in the qubit loop, the difference being about one flux quantum $\Phi_0$.  The critical current of the measurement SQUID is sensitive to this difference in flux, allowing us to discriminate between the two qubit states by ramping the SQUID bias and measuring the current when the SQUID switches into the voltage state.

This sensitivity to qubit flux is only necessary during measurement, and is in fact detrimental during qubit operation.  If the SQUID is sensitive to flux from the qubit, then the qubit is also sensitive to flux from the SQUID; noise and dissipation in the SQUID circuit---in particular from the shunt resistance $R_s$---will decohere the qubit state.  We would like to be able to modulate the SQUID's flux sensitivity, turning off the coupling during qubit operation, and turning it on only for measurement.  The three-junction design makes this possible.

When bias current $I_1=I_{sq}$ is applied to the SQUID, it divides into the upper and lower branches of the loop.  The lower branch has a single Josephson junction with critical current $I_0$, whereas the upper branch has two larger Josephson junctions each with critical current $\alpha I_0$.  The total current is $I_{sq} = I_U + I_L = \alpha I_0\sin(\delta/2) + I_0\sin(\delta)$, where $\delta$ is the superconducting phase difference across the loop.  The circulating current in the loop is $I_{circ} = I_U - I_L = \alpha I_0\sin(\delta/2) - I_0\sin(\delta)$.  This circulating current couples via a fixed mutual inductance $M_{sq}/2$ in each branch to the qubit loop, causing current $I_2=I_q=(M_{sq}/2L_q)I_{circ}$ to flow.

\begin{figure}[b]
\begin{center}
\includegraphics[trim=0 1.7in 0 0.2in] {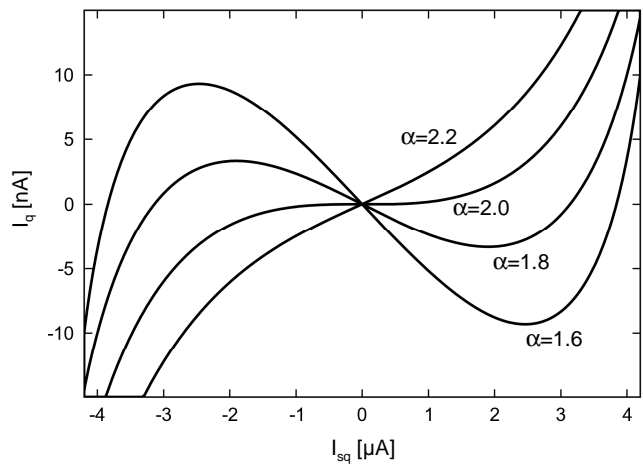}
\end{center}
\caption{Theoretical transfer function of the 3-junction SQUID.  The plot shows the induced qubit current $I_q$ versus SQUID bias current $I_{sq}$, for four values of the junction size ratio $\alpha$, with device parameters as for Fig. \ref{fig:Layout}.  At points where $dI_q/dI_{sq} = 0$, the qubit will be insensitive to noise and dissipation from the SQUID.  The design value of $\alpha = 1.7$ ensures the existence of an insensitive point that is not too close to the critical current of the SQUID.  The physical origin of decoupling is easily understood for $\alpha=2$.  At the bias $I_{sq} \approx 0$ the Josephson inductances of the upper and lower branches are equal, producing a symmetric flow of current and no net flux to the qubit.}
\label{fig:SquidTheory}
\end{figure}

A plot of $I_{sq}$ versus $I_q$ is shown in Fig. \ref{fig:SquidTheory} for four values of $\alpha$.  The SQUID and qubit are decoupled at points of zero slope; these ``insensitive points'' exist for $\alpha \leq 2$.  Away from the insensitive point, the inductances become unbalanced and the transfer function has nonzero slope, so that SQUID and qubit are again coupled.  When measuring the qubit, we ramp $I_{sq}$ toward the critical current, turning the coupling on and allowing the SQUID to discriminate between the tunneled and non-tunneled qubit states.  Because of unavoidable variations in junction size during fabrication, we typically design for $\alpha \approx 1.7$ to ensure that an insensitive point will exist, yet not be too close to the critical current of the SQUID.

\begin{figure}[tbp]
\begin{center}
\includegraphics[trim=0 0.7in 0 0.1in] {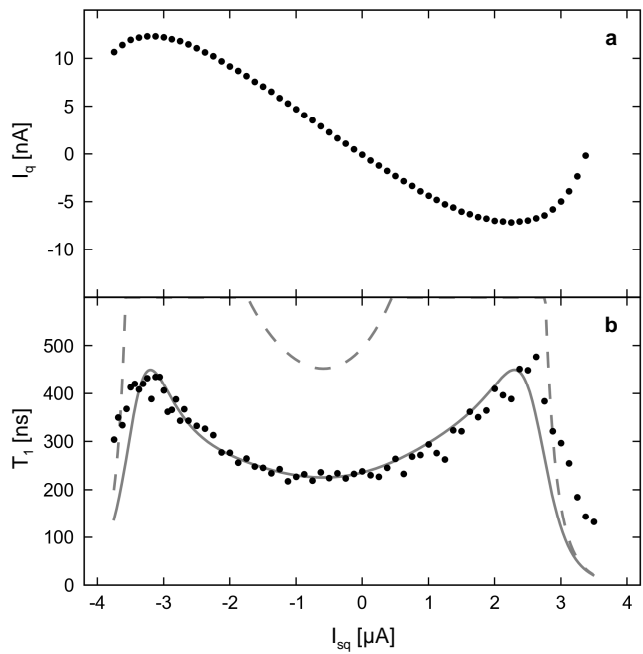}
\end{center}
\caption{Measured SQUID transfer function and its effect on qubit lifetime.  (\textbf{a}) Qubit current $I_q$ versus SQUID bias $I_{sq}$, measured as described in the text.  Outside the range shown, the SQUID switches prematurely, preventing reliable qubit operation.  (\textbf{b}) Measured qubit lifetime $T_1$ (dots) along with theoretical curves.  The dashed line is the prediction from the transfer function alone, while the solid line adds to this a constant dissipation corresponding to a lifetime of $450\,~\textrm{ns}$.  Best fit is for $R_s = 11\,~\Omega$.}
\label{fig:T1Data}
\end{figure}

The tunability of the phase qubit with flux allows us to measure the transfer function of the SQUID.  We first set $I_{sq} = 0$ and find the qubit resonance frequency with spectroscopy \cite{Simmonds2004}.  When the SQUID bias $I_{sq}$ is set to a new value, the circulating current in the SQUID produces an offset flux $\Delta\Phi_{sq} = L_q I_q$ in the qubit, shifting its frequency.  We then adjust the flux bias $\Delta\Phi_{fb} = M_{fq}I_{fb}$ to bring the qubit frequency back to its original value.  For the qubit frequency to be unchanged, these two fluxes must cancel, and we have $I_q = -(M_{fq}/L_q) I_{fb}$.  By repeating this procedure for a range of values of SQUID bias, we build up a measurement of the transfer function $I_q(I_{sq})$, as shown in Fig. \ref{fig:T1Data}a.  An alternative method is to hold constant the $\ket{0}$-state tunneling rate instead of the resonance frequency, but we found that the resonance frequency was a more sensitive probe of the qubit current $I_q$ and more immune to systematic errors.

Next we measure $T_1$ as a function of SQUID bias by applying a $\pi$-pulse to the qubit and measuring the decay of the $\ket{1}$-state probability with time.  As in the measurement of the transfer function, a flux offset is applied at each SQUID bias to keep the qubit frequency constant, removing any frequency dependence of the dissipation. Figure \ref{fig:T1Data}b shows the measured $T_1$ data along with predictions from the impedance-transformer model using the measured transfer function.  The lifetime $T_1$ varies with SQUID bias as expected, increasing as the transfer function flattens and reaching its maxima at the insensitive points $dI_q/dI_{sq}=0$.  Beyond these biases, the transfer function has a large derivative and $T_1$ drops sharply \cite{Fitting}.

A full prediction of $T_1$ must add a parallel dissipation channel to account for decay from other dissipation mechanisms, especially at the insensitive point where there is no dissipation from the transformer.  Taking the maximum observed value of $450\,\textrm{ns}$, consistent with dielectric loss due to the a-Si:H dielectric of the device \cite{Martinis2005}, we find excellent agreement between the data and theory.  Using the measured slope $dI_q/dI_{sq}$, we find best agreement with shunt resistance $R_s=11\,\Omega$ transformed by the SQUID.

While the agreement between theory and experiment is encouraging, there are two simplifying assumptions in the impedance transformation model that merit discussion.  First, the model assumes that the coupling element is a purely inductive circuit, which has a frequency-independent transfer function.  In the actual circuit, the SQUID capacitance leads to frequency-dependent effects that drastically alter the transfer function near the self-resonant frequency of the SQUID.  In our device the SQUID self-resonance frequency is $\sim 15\,\textrm{GHz}$, well above the qubit frequency of $6.75\,\textrm{GHz}$.  Numerical calculations have shown that at $6.75\,\textrm{GHz}$, the SQUID capacitance simply increases the transfer function $\left|dI_2/dI_1\right|^2$ by a factor of $\sim 2$.  In the circuit, the resistance from the $30\,\Omega$ shunt resistor in parallel with the $50\,\Omega$ bias line is effectively modified by this effect to give an effective shunt resistance of $(30\,\Omega || 50\,\Omega)/2\approx 9\,\Omega$.  This agrees well with the best fit value of the shunt resistance $11\,\Omega$.

We note that the impedance transformation measured here corresponds to $|dI_1/dI_2|^2 \sim 10^5$, and we have confirmed the magnitude of this transformation to better than a factor of 2.

Secondly, the simple transformer theory predicts a diverging impedance $1/\textrm{Re}\,Y_2$ at the extrema of the current transfer function, where $dI_q/dI_{sq} = 0$.  In the experiment, we expect divergences to be rounded off by higher-order processes.  The second-order effect can be calculated straightforwardly as follows: the shunt resistor $R_s$ produces a quantum noise current \cite{Schoelkopf2003,Martinis2003} with a one-sided spectral density given in the limit $T \to 0$ by $S_{I_1}(f) = 2hf/R_s$ for $f > 0$, and $S_{I_1}(f) = 0$ for $f < 0$,
where $h$ is Planck's constant.  The (complex) noise current $I_1(t)$ produced by this resistor is transformed by the coupler to a noise current at port 2 with Taylor expansion $I_2(t) = \textrm{const.} + (dI_2/dI_1) I_1(t) + (d^{2}I_2/dI_1^{2})I_1(t)^{2}/2$.  We calculate the spectral density of this transformed current by inserting the Fourier transform of $I_1(t)$, assuming random phases of all the frequency components.  This gives for the transformed spectral density
\begin{eqnarray}
S_{I_2}(f) & = & \left|\frac{dI_2}{dI_1}\right|^{2}S_{I_1}(f) \nonumber \\
           & + & \frac{1}{2!}\left|\frac{d^{2}I_2}{dI_1^{2}}\right|^{2} \int_{0}^{f} df'S_{I_1}(f')S_{I_1}(f-f').
\label{eqn:xformspec}
\end{eqnarray}
The first term in this spectral density corresponds to the simple linear impedance transformation model discussed previously.  The second term corresponds to dissipation due to (nonlinear) downconversion, in which the photon from port 2 at frequency $f$ is converted to two photons at frequencies $f'$ and $f''$ that are absorbed by the environment at port 1, where $f' + f'' = f$.  The second-order process will dominate at the extrema of the transfer function where $dI_2/dI_1=0$, and lead to a finite lifetime.

The transformed spectral density may be evaluated for the spectral density of a resistor, giving
\begin{equation}
\textrm{Re}\,Y_2 = \frac{S_{I_2}(f)}{2hf} = \frac{1}{R_s}\left(\left|\frac{dI_2}{dI_1}\right|^{2}+\frac{1}{6}\left|\frac{d^{2}I_2}{dI_1^{2}}\right|^{2}\frac{hf^{2}}{R_s}\right).
\end{equation}
This calculation can be extended to arbitrary order by keeping terms in the Taylor expansion of the transfer function.  For a resistor the integrals can be evaluated exactly yielding
\begin{equation}
\textrm{Re}\,Y_2 = \frac{1}{R_s}\sum_{k=1}^\infty \frac{1}{k!(2k-1)!} \left|\frac{d^{k}I_2}{dI_1^{k}}\right|^{2} \left( \frac{2hf^2}{R_s} \right)^{k-1}.
\end{equation}
The predicted lifetime due to the second-order process is $\sim 100\,\mu\textrm{s}$, far from the limiting value of the lifetime observed in our experiment.  Thus, even accounting for second-order noise processes, we find that the SQUID is completely decoupled, as desired.  The lifetime is thereby limited to the observed $450\,~\textrm{ns}$ due to another loss mechanism, most likely dielectric loss.

To illustrate the utility of this transformer theory, we consider the case of dissipation arising from the microwave lines used to control the qubit.  Although coupling of microwaves to the qubit is set by a coupling capacitor or mutual inductance, in a real experimental device there is typically some uncertainty in the exact coupling strength and resulting dissipation due to, for example, complex microwave modes.  If the strength of the microwave coupling is simply measured by knowing the strength of the microwave amplitude driving the chip and the Rabi oscillation frequency, then the current transformation can be determined for the real physical coupling element.  When the measured $T_1$ is compared to predictions from an impedance transformation calculated with this theory, one can determine whether coupling via this environmental mode dominates the observed decay.  We emphasize this theory allows comparison to the actual experimental system, not just an idealized circuit model.

In conclusion, we have directly measured in a Josephson phase qubit the current transfer function of a tunable 3-junction SQUID and its transformed dissipation.  The variation in qubit lifetime as a function of SQUID bias was analyzed with a simple model based on the classical impedance transformation from the measured SQUID-qubit transfer function.  The dissipation predicted by this model agrees quantitatively with measurements for this non-linear coupling element.  As more sophisticated quantum circuits are developed---for example in implementing tunable coupling---having a simple method to calculate environmental dissipation will become increasingly important.

Devices were made at the UCSB and Cornell Nanofabrication Facilities, a part of the NSF-funded National Nanotechnology Infrastructure Network.  This work was supported by DTO under grant W911NF-04-1-0204 and NSF under grant CCF-0507227.

\bibliographystyle{apsrev}

\begin{thebibliography}{3}
\bibitem{Fitting} In Fig. 4, the small deviation from theory at currents above the insensitive point may be due to small systematic errors in determining the current transfer function.
\end{thebibliography}

\end{document}